\documentclass[11pt,a4paper]{article}
\usepackage{jheppub}

\usepackage[T1]{fontenc}
\usepackage{lmodern}

\newcommand{\ee}{\mathrm{e}}
\newcommand{\im}{\mathrm{i}}
\newcommand{\de}{\mathrm{d}}
\newcommand{\db}{\boldsymbol{\de}}
\makeatletter
\let\Re\@undefined
\let\Im\@undefined
\makeatother
\DeclareMathOperator{\Re}{Re}
\DeclareMathOperator{\Im}{Im}
\newcommand{\vect}[1]{\mathbf{#1}}
\newcommand{\spc}{{\,}}
\newcommand{\sep}{{\qquad}}

\DeclareMathOperator{\Hodge}{\star}
\newcommand{\Iprod}[2]{\langle {#1}, {#2} \rangle}

\preprint{\parbox[t]{.25\textwidth}{\raggedleft DFPD-2012-TH-18\\IFIC/12-73\\WITS-CTP-107}}

\title{On anharmonic stabilisation equations for black holes}

\author[a,b]{Pietro Galli,}

\affiliation[a]{Departament de F\'{\i}sica Te\`{o}rica and IFIC (CSIC-UVEG)\\
Universitat de Val\`{e}ncia\\
C/ Dr.~Moliner, 50, 46100 Burjassot (Val\`{e}ncia), Spain}

\emailAdd{Pietro.Galli [at] ific.uv.es}

\author[b]{Kevin Goldstein}

\affiliation[b]{National Institute for Theoretical Physics (NITheP)\\
School of Physics and Centre for Theoretical Physics\\
University of the Witwatersrand\\
WITS 2050, Johannesburg, South Africa}

\emailAdd{Kevin.Goldstein [at] wits.ac.za}

\author[c]{and Jan Perz}

\affiliation[c]{Istituto Nazionale di Fisica Nucleare\\
Sezione di Padova\\
Via Marzolo, 8, 35131 Padova, Italy}

\emailAdd{Jan.Perz [at] pd.infn.it}

\abstract{We investigate the stabilisation equations for sufficiently
  general, yet regular, extremal (supersymmetric and
  non-supersymmetric) and non-extremal black holes in four-dimensional
  $N=2$ supergravity using both the H-FGK approach and a
  generalisation of Denef's formalism.  By an explicit calculation we
  demonstrate that the equations necessarily contain an anharmonic
  part, even in the static, spherically symmetric and asymptotically
  flat case.}


\begin{document}

\maketitle

\section{Introduction}

Among the efforts to systematise the construction of non-supersymmetric black hole solutions in four-dimensional $N=2$ supergravity one can discern two intersecting lines of research: on the one hand the generalisation \cite{Galli:2009bj,Galli:2010mg} of Denef's formalism \cite{Denef:2000nb}, applicable to stationary extremal black holes, and the H-FGK approach \cite{Mohaupt:2011aa,Meessen:2011aa} for static extremal and non-extremal solutions on the other. In distinct ways each arrives at a set of relationships, which we shall call stabilisation equations, between duality-covariant combinations of physical degrees of freedom and ans\"atze for spatial functions $H^M\!(\vect{x})$. These relationships remain unchanged for various types of black holes, which means that all black hole solutions (supersymmetric, extremal, non-extremal) in a given model take the same form in terms of the functions $H$ and only the functions themselves vary.

For supersymmetric extremal solutions, functions $H$ are known to be harmonic, with poles corresponding to physical magnetic and electric charges carried by the black hole \cite{Sabra:1997kq,Behrndt:1997ny}. In the context of the H-FGK formalism a harmonic ansatz has been used also for non-supersymmetric, static, spherically symmetric extremal black holes, whereas for their non-extremal counterparts a hyperbolic (exponential) ansatz has been employed \cite{Mohaupt:2010fk,Galli:2011fq,Meessen:2012su,Bueno:2012jc,GOPS:2012}.

In this short note we examine the exhaustiveness of these ans\"atze in the static, spherically symmetric case, i.e.~with the metric of the form
\begin{equation}
\label{eq:StaticMetric}
\de s^2 = -\ee^{2U(\tau)}\de t^2 + \ee^{-2U(\tau)}\left(\frac{r_0^4}{\sinh^4(r_0\tau)}\de\tau^2 + \frac{r_0^2}{\sinh^2(r_0\tau)}(\de\theta^2 + \sin^2\!\theta\,\de\phi^2)\right),
\end{equation}
providing in the process some portions of a dictionary between the generalised formalism of Denef and the H-FGK formulae.

\section{Non-superysmmetric extremal black holes}

In \cite{Galli:2010mg}, to which we refer the reader for the description of the general setup and whose numerical conventions we follow (occasionally adopting some of the notation from the H-FGK literature), the generating single-center underrotating solution \cite{Bena:2009ev} for the metric warp factor $U(\vect{x})$ and the complex scalars $z^a(\vect{x})$ from $n_\mathrm{v}$ vector multiplets in models with cubic prepotentials has been recast in the form of stabilisation equations
\begin{equation}
2\Im\left(\ee^{-U-\im\alpha}\,\Omega^M\!(z,\bar{z})\right) = H^M\!(\vect{x}) \spc,
\end{equation}
where $\Omega(z,\bar{z})$ is the covariantly holomorphic symplectic section (period vector) of special geometry, $\alpha$ is a phase and the single superscript $M$ is understood to run over $2(n_\mathrm{v}+1)$ components, otherwise indexed with subscripts and superscripts ${}^{0,a}$ and $_{0,a}$. $H$ was written in \cite{Galli:2010mg} as a sum of harmonic functions and a ratio of harmonic functions (note the minus sign in the zeroth magnetic component):
\begin{equation}
\label{eq:HwithJ}
(H^M) = \left(h^0-p^0\tau,0;0,h_a+q_a\tau\right) + \left(0,0;\frac{b+J\tau^2\cos\theta}{h^0-p^0\tau},0\right),
\end{equation}
where $\tau$ is a radial coordinate and the anharmonic part persists also in the absence of rotation ($J = 0$), when the solution reduces to that of \cite{Cardoso:2007ky,Gimon:2007mh}. Although the quotient form of $H$ was later confirmed by \cite{Bossard:2012xs}, one could nonetheless wonder whether the anharmonic part is necessary (as opposed to being an artifact of the specific rewriting with the particular coefficients used) and whether the solutions that seem to require it do not carry NUT charge (which would render them only locally asymptotically flat).

To answer these questions we solve the spherically symmetric, static case of the $t^3$ model\footnote{Normalization: $\Omega_0 = \frac{5}{6}(\Omega^1)^3/(\Omega^0)^2$. (Here, unlike in \cite{Galli:2010mg}, $\Omega_0$ stands for one of the components of $\Omega$.)} for the charge configuration $(Q^M) = (0,p^1;q_0,0)$, dual to that in eq.~\eqref{eq:HwithJ} for $n_\mathrm{v} = 1$. It is easiest to start with the equation (2.27) of \cite{Meessen:2011aa},\footnote{This equation can also be derived from a further generalisation of Denef's formalism to non-extremal solutions. Although this derivation does not appear in the literature, we do not include the rather technical details here since they are not directly relevant to our discussion.} which corresponds to the equation of motion for the warp factor:
\begin{equation}
\label{eq:EOMforU}
\frac{1}{2}\frac{\partial\log\ee^{-2U}}{\partial H^M}(\ddot H^M - r_0^2 H^M) + \left(\frac{\dot H^M H_M}{2\ee^{-2U}}\right)^2 = 0 \spc,
\end{equation}
where the dot denotes differentiation with respect to $\tau$, the index $M$ has been lowered with the symplectic form $\left(\begin{smallmatrix}0 & -1\\1 & 0\end{smallmatrix}\right)$ and where
\begin{equation}
\ee^{-2U} = \sqrt{-\tfrac{10}{3}(H^1)^3 H_0 - (H^0 H_0)^2 - 2 H^0 H^1 H_0 H_1 + \tfrac13(H^1 H_1)^2 + \tfrac{8}{45} H^0 (H_1)^3}\spc.
\end{equation}
As remarked in \cite{Meessen:2011aa}, when $r_0 = 0$ (extremal black holes) and upon assuming,
\begin{equation}
\label{eq:HdotH}
\dot H^M H_M = 0\spc, 
\end{equation}
(\ref{eq:EOMforU}) reduces to
\begin{equation}
\label{eq:simpleEOM}
\frac{\partial\log\ee^{-2U}}{\partial H^M}\ddot H^M = 0\spc,
\end{equation}
which can be solved by harmonic functions $\ddot H^M = 0$. The harmonic function solution sets each term in (\ref{eq:simpleEOM}) to zero individually.

One may however relax the assumption (\ref{eq:HdotH}), setting $H_1 = 0$, taking only the two functions corresponding to non-vanishing charges to be harmonic with arbitrary coefficients ($H^1 = A^1 + B^1\tau$, $H_0 = A_0 + B_0\tau$) and leaving $H^0$ unspecified. Eq.~\eqref{eq:EOMforU} then becomes 
\begin{equation}
(A_0 + B_0\tau)^2 H^0\ddot H^0 - \frac12\left(B_0 H^0 - (A_0 + B_0\tau)\dot H^0\right)^2=0\spc,
\end{equation} 
a model-dependent differential equation for $H^0(\tau)$, whose solution reads
\begin{equation}
\label{eq:H0ext}
H^0 = \pm\left(c_1\sqrt{A_0 + B_0\tau} + \frac{c_2}{\sqrt{A_0 + B_0\tau}}\right)^2\spc,
\end{equation}
with constants of integration $c_1, c_2$. The remaining equations of motion fix the coefficients as either
\begin{equation}	
c_1 = 0\spc, \sep B_0 = -q_0\spc, \sep B^1 = p^1\spc,
\end{equation}
in exact analogy with eq.~\eqref{eq:HwithJ}, or
\begin{equation}
c_1 = 0\spc, \sep c_2 = 0\spc, \sep B_0 = 0\spc, \sep B^1 = 0\spc,
\end{equation}
which leads to a (doubly extremal) solution with constant scalars. The other parameters and the overall sign in \eqref{eq:H0ext} are determined by the asymptotic boundary conditions. In particular, for the non-constant solution (we suppress the superscript $1$ on the single scalar $z = \Omega^1/\Omega^0$):
\begin{equation}
\operatorname{sgn}(H^0) = -\operatorname{sgn}(\Re z_\infty)\spc, \sep
c_2^2 = \left\lvert\frac{\Re z_\infty}{\Im z_\infty}\right\rvert.
\end{equation}

\section{Non-extremal black holes}

For $r_0 \neq 0$ and with the additional assumption $\dot H^M H_M = 0$, eq.~\eqref{eq:EOMforU} reduces to
\begin{equation}
\frac{\partial\log\ee^{-2U}}{\partial H^M}(\ddot H^M - r_0^2 H^M) = 0\spc,
\end{equation}
which can be solved by hyperbolic functions $\ddot H^M = r_0^2 H^M$. Searching for a more general solution we take, similarly to the extremal case above, $H_1 = 0$, $H^1 = A^1\cosh(r_0\tau) + \frac{B^1}{r_0}\sinh(r_0\tau)$ and $H_0 = A_0\cosh(r_0\tau) + \frac{B_0}{r_0}\sinh(r_0\tau)$. $H^0$ is then determined from 
\begin{equation}
\begin{split}
&\left(A_0\cosh(r_0\tau) + \tfrac{B_0}{r_0}\sinh(r_0\tau)\right)^2 H^0 (\ddot H^0 - r_0^2 H^0)\\
&-\frac12\left[\Bigl(r_0 A_0\sinh(r_0\tau) + B_0\cosh(r_0\tau)\Bigr) H^0 - \left(A_0\cosh(r_0\tau) + \tfrac{B_0}{r_0}\sinh(r_0\tau)\right)\dot H^0\right]^2=0\raisetag{10ex}
\end{split}
\end{equation}
and turns out to be
\begin{equation}
H^0 = \pm\frac{\left(c_1\cosh(r_0\tau) + \frac{c_2}{r_0}\sinh(r_0\tau)\right)^2}{A_0\cosh(r_0\tau)+\frac{B_0}{r_0}\sinh(r_0\tau)}\spc.
\end{equation}
Numerical tests indicate that the analytical solution for the coefficients
\begin{gather}
B_0 = c_2 A_0\spc, \sep B^1 = c_2 A^1\spc,\\
\begin{split}
& c_2 = \pm c_1\left( 75 (A^1)^4 (A_0)^2 (p^1)^2 - 45 c_1^4 A^1 A_0 (p^1)^2 + 45 c_1^4 (A^1)^2 p^1 q_0 + 25 (A^1)^6 (q_0)^2 + 9 c_1^8 r_0^2\right.\\
&\left.{} + 60 (A^1)^3 A_0 c_1^4 r_0^2 + 100 (A^1)^6 (A_0)^2 r_0^2\right)^\frac{1}{2}\Big/\left(4 c_1^4+10 A_1^3 A_0\right)\raisetag{4ex}
\end{split}
\end{gather}
is the only admissible solution. Such coefficients lead to a constant scalar, which must take the extremal attractor value. It follows that $c_1=0$, so ultimately $H^0 = 0$, the solution is given purely in terms of hyperbolic functions (and compatible with the condition $\dot H^M H_M = 0$).

\section{Discussion and conclusions}

In spite of their different origins, the non-supersymmetric extension of Denef's approach and the H-FGK formalism both match the scalar degrees of freedom with the vector part of the action in the same way, one that respects duality covariance. The corresponding non-differential stabilisation equations have consequently (up to the differences in conventions) identical form. The fact that the $H$-functions differ stems from the specific additional assumptions made in the H-FGK literature, namely that $\dot H^M H_M = 0$ and that the rest of eq.~\eqref{eq:EOMforU} vanishes term by term.

The condition $\dot H^M H_M = 0$ in the BPS context is synonymous with the absence of NUT charge \cite{Bellorin:2006xr}. For the non-supersymmetric extremal solution discussed here this cannot be the case, since all the equations of motion are satisfied with the static metric \eqref{eq:StaticMetric}, whose NUT charge is $0$. Indeed, ref.~\cite{Galli:2010mg}, eq.~(3.28) showed that the spatial Hodge dual of the spatial exterior derivative of the one-form $\omega$ encoding the relevant part of the metric depends on two terms,
\begin{equation}
\boldsymbol{\Hodge_0}\db\omega = \Iprod{\db H}{H} - 2\ee^{-2U}\boldsymbol{\eta}\spc,
\end{equation}
the first of which directly generalises $\dot H^M H_M$. (The second term measures the non-closure of the fake electromagnetic field strength two-form introduced therein.) We see that for the left-hand side to be zero it suffices that rather than each part vanishes, as happens for BPS solutions, the two terms only cancel each other, as in the extremal example discussed above.\footnote{Cf.~also \cite{GMO:2012} for the discussion of gauge dependence of the condition $\dot H^M H_M = 0$.}

It is worth pointing out that the inverse harmonic part of the functions $H$ is essential for the non-trivial behaviour of the real parts of $z^a$, usually referred to as axions. We have checked that the constant $c_2$ (or $c$ in \cite{Cardoso:2007ky}, $B$ in \cite{Gimon:2007mh} and $b$ in eq.~\eqref{eq:HwithJ}), originating here from the product $H^0 H_0$, cannot be consistently extracted from the other constants when $H^M$ are purely harmonic (the system equations that one would write does not admit any solution), even if none of them were a priori vanishing.

The non-extremal case remains less lucid. The existence of non-hyperbolic solutions has been postulated in \cite{Mohaupt:2012tu}, but the non-hyperbolic part of the natural generalization of the extremal anharmonic solution in our example turned out to be zero. Arguably however, by setting some of the $H^M$ to be harmonic or hyperbolic functions we might not yet have searched for the most general extremal or non-extremal solution.

\acknowledgments

We greatly benefitted from discussions with Prof.~P.~Meessen and Prof.~T.~Ort\'{\i}n. PG wishes to thank the University of the Witwatersrand for hospitality. The work of PG has been supported in part by grants FIS2008-06078-C03-02 and FIS2011-29813-C02-02 of Ministerio de Ciencia e Innovaci\'on (Spain) and ACOMP/2010/213 from Generalitat Valenciana. The work of KG is supported in part by the National Research Foundation. Any opinion, findings and conclusions or recommendations expressed in this material are those of the authors and therefore the NRF do not accept any liability with regard thereto. JP's work has been supported by the ERC Advanced Grant no.~226455 (SUPERFIELDS).

\appendix
\section{Comparison of conventions}

Some of the original symbols have been replaced with those used here to make the meaning of the expressions clearer. Comparison with the respective papers provides a dictionary. $\hat\Omega = \ee^{-U-\im\alpha}\Omega(z,\bar{z})$.
\begin{center}
\begin{tabular}{l c c c}
\hline
 & ref.~\cite{Galli:2010mg} & ref.~\cite{Meessen:2011aa} (H-FGK) & here \\
\hline
metric signature & $(-,+,+,+)$ & $(+,-,-,-)$ & $(-,+,+,+)$ \\
$\tau\in$ & $(0,\infty)$ & $(0,-\infty)$ & $(0,\infty)$ \\
physical scalars & $z^a$ & $Z^i$ & $z^a$\\
vector super- and subscript & $I = 0,a$ & $\Sigma = 0,i$ & not used\\
single index & not used & $M = {}^\Sigma,{}_\Sigma$ & $M$\\
$H$-functions & $2\Im\hat\Omega = \mathcal{J}$ & $\Im\hat\Omega^M = H^M$ & $2\Im\hat\Omega^M = H^M$ \\
symplectic form &
$\left(\begin{smallmatrix}0 & -1\\1 & 0\end{smallmatrix}\right)$ &
$\left(\begin{smallmatrix}0 & 1\\-1 & 0\end{smallmatrix}\right)$ &
$\left(\begin{smallmatrix}0 & -1\\1 & 0\end{smallmatrix}\right)$ \\
warp factor & $\ee^{-2U} = \im\hat\Omega^M\bar{\hat\Omega}_M$ & $\ee^{-2U} = -\frac{\im}{2}\hat\Omega^M\bar{\hat\Omega}_M$ & $\ee^{-2U} = \im\hat\Omega^M\bar{\hat\Omega}_M$\\
poles of BPS $H$ & $\Gamma\tau$ & $-\frac{Q^M}{\sqrt{2}}\tau$ & $Q^M\tau$ \\
\hline
\end{tabular}
\end{center}
Note the symplectic form hidden in the expression for the warp factor.

In this paper by ``stabilisation equations'' we mean $\Im\hat\Omega \propto H$, whereas the H-FGK papers use that term for the relations between the real and imaginary parts of $\hat\Omega$: $\Re\hat\Omega = \Re\hat\Omega(\Im\hat\Omega)$.

\bibliographystyle{JHEP}
\bibliography{Anharmonic}

\end{document}